# The Software Garden

Federico D. Sacerdoti
*AienTech*, *New York, NY*

**Abstract**

This paper describes a practical method of developing custom HPC software products using a store of libraries and tools independent from the OS called a "garden". All dependencies from the product to libraries of the underlying OS distribution are carefully severed, isolating the package from instability due to system upgrades and ensuring repeatable deterministic builds on different flavors of Linux. The method also guarantees multiple versions of a software product may exist together and function correctly, greatly facilitating upgrade and rollback. The method is the first known system to expose all past software versions to the designer, and support deterministic single-package rollback without affecting other installed software. An application of this method for building a high performance trading system in C++ is presented.

## 1.  Introduction

A problem facing DevOps and system administrators is to provide a software product with the system libraries it requires to run correctly. This task is complicated by the need to quickly apply OS and security upgrades, possibly unilaterally, i.e. without the involvement of software developers. It is often unclear if production software will be affected by an upgrade given the ability of popular tools like *yum* and *apt* to upgrade many packages in a single transaction. Furthermore required upgrades are frequent, and prone to failure [4].

We provide a method of constructing software for HPC environments that ensures runtime dependencies change only when expected, i.e. during the development cycle, not during dead-of-night upgrades. This method has no speed penalty, and allows new compilers and libraries to be tested and used in production without requiring a "forklift upgrade" that affects all versions of the product.

A related problem is build stability: how to ensure a software product is built with the correct compiler and libraries even after those tools are upgraded on the OS. True stability produces identical results when the product is built on different of Linux distributions, or even slightly different versions of the same distribution, a common situation in HPC. Specifically how can the compiler toolchain, libraries, and header files be organized such that a) the same version is present for all developers and production, b) they can correctly build previous versions of software that were made and tested in the past, and c) new compiler and library versions may be tried without affecting old product versions. The latter two are most difficult for Ubuntu/Redhat-style Linux platforms that fundamentally support only a single version of a package. Goal (b) is important for reliable rollback by DevOps, and for bug reproduction and repair by developers; goal (c) greatly aids development and software maintenance.

Our method of building software, called the Software Garden, addresses both problems by attacking the issue of multiple versions head on and early. No other system we are aware of exposes a full array of multiple software versions to the designer, and therefore falls short of these goals. NIX [1] comes closest; it supports multiple versions of a package, but is designed as an OS replacement, and hides all but one version from the user[1]. The Garden is a development system; it operates on source files and supports incremental development, package testing, and sophisticated composition hints for a package's inclusion in other software. NIX is a useful packaging system that takes a set of source tarballs as input and provides a complete OS distribution as output [2]. The Garden uses NIX's backend engine to name package versions and manage dependencies, and leverages all NIX-packaged software.

The Garden's aim is to empower software designers to deliver software that *remains stable forever*. We not provide a traditional OS distribution with a single version of /bin. The Garden's output does not look familiar, and its use is far different. No "default version" is assigned to a user, as such a thing is not stable. In fact the Garden's aim is to obliterate the traditional notion that /bin and /lib and /usr/include contain "some thing" at all; they are a mist that can and will change. Therefore Garden software echews them, and provides no easy handle to a single "blessed" version. The version to use is exactly the one that was tested with a client's logic, and no more. We will show this idealism has some important exceptions, but fundamentally the Garden's goal is to coexist but be completely independent from the OS.

In this paper we illustrate the novel mechanisms of the Garden through a real-world example of a software product and its supporting ecosystem.

---

[1] A NIX user is given an "environment generation" with symlinks to one version of each package in a per-user *bin/* directory.

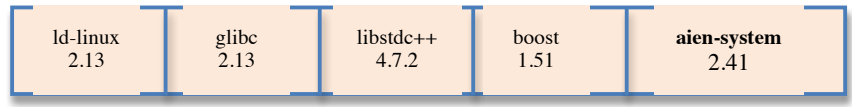

Figure 1) Version vector of a software package. The *aien-system* package requires the other software packages in the vector, and was tested with those versions specifically. A rollback to a previous version of *aien-system* is a vector rollback. The system must support multiple installed versions because other running code, which we cannot roll back, may rely on something in the original vector.

## 2. Background

When Anderson and Patterson define stages of software installation [3], they describe *packaging* and *merging*. Packaging binds pathnames to an application version; merging resolves conflicts between multiple versions, generally by linking them to a common directory.

The authors state supporting multiple package versions simultaneously is both difficult and unsolved. However it is of critical importance. A software version number is more accurately described as a vector of version numbers: its own and that of each dependency (Figure 1). Without multiple version support, the system is forcing a promise that all software uses the same vector except for one element, its own. This rigidity may lead to breakage: the designer has carefully tested a package with one vector, but an upgrade forces another vector. Without multiple simultaneous versions we have limited options: either delay or abandon upgrades, or re-test all software packages with the upgraded package[2]. The former stifles progress and prevents important security updates, the latter is impractical and error prone.

### 2.1. CASE STUDY

A small team was assembled to build an HPC message-passing product. C++ was chosen as the language, GIT for source control [6], WAF for make [7]. Work proceeded using standard practices, with developers using Ubuntu 10.X workstation, and the gcc 4.6.x compiler. Within weeks the first breakage occurred. One developer upgraded to Ubuntu 11.10 and over the next several days unwittingly introduced C++ constructs only supported by the new gcc, which he was not aware had been upgraded. The knife slips in easily: "for me it just works" was the report. Similar issues followed for Python components of the product.

One solution would have been to enforce one OS version for all developers, all DevOps, and all production systems. Unfortunately this is not a practical solution. Developers require a Workstation distribution with graphical support; production systems use a similar but not identical Server distribution. It also stifles adoption of new tools: the new C++ constructs helped that developer save time and avoid bugs; but the team is too busy to forklift upgrade the compiler on all systems, which is difficult to test and execute. Finally, once such an upgrade occurs, rollback to a pre-upgrade version is difficult and error prone.

The CDE system [8] was considered but the speed degradation from running binaries permanently under strace was not tolerable. All development and production was switched to the Garden system, a process that took only a few days due to the large amount of cleanly-built software made available by NIX. After the switch all developers had a single version of gcc regardless which OS version they run (Figure 2). They were free to try new versions of gcc without breaking the production build. The developers and DevOps quickly become reliant on the Garden, and requested other products be moved into it.

### 2.2. OTHER METHODS

Other techniques to address these issues include statically linking all executables, which solves the goals (a) and (c) and the vector-version problem. However it uses more resources: disk space, and memory on multi-core machines as running copies of the same executable and common libraries are unshared [10]. It slows down the development cycle due to the required full-relink. Finally static linking does not cover scripts, executables with plugins, or interpreted languages, and does not address the goal of repeatable and stable builds (a,b). Despite these shortcomings, it is often used where possible.

Occasionally multiple Virtual Machine images are used, one for each software configuration. This solution requires more resources than static linking, as there are as many software configurations as unique vector-versions, and not even the OS distribution is shared. Moreover supporting rollback of a single package is not possible, as only one version of a library exists in each VM image.

The GLIBC library employs the complex and rarely used versioning technique that requires the library to act like all previous versions of itself when asked [11]. This technique imposes significant development and testing overhead. The commercial RPath offering based on Conary uses source control techniques to advance and rollback sets of versioned software packages, but only supports one installed version at a time [12]. Per-package versioning is often the only available solution for dynamically-linked base libraries, leaving users with commands like `python2.7` [9]. This solves neither the isolation problem nor the stable build problem.

---

[2] Carefully crafting a .spec file (RPM) with detailed *requires* elements forces exactly this choice: a freeze, or a full re-test.

```
# Treetop
gcc    = "bj61jjvy9fnm3xxpyq12zpwx2mgg09c8-gcc-4.6.1";
glibc  = "vxycdbhcj720hzkw2px7s7kr724-glibc-2.12.2";
python = "vzpvrymybznxha6hadj0ww68vx-python-2.7.1";
stdenv_linux = "n4yy8dn2gfa2vnifdjz0b-stdenv-linux";
json   = "7r0f0jjh128ps51l81-cajun-2.2";
export = [ gcc python stdenv_linux ];
```

```
$ garden avail gcc
/nix/store/bj61jjvy9fnm3xxpyq12zpwx2mgg09c8-gcc-4.6.1
/nix/store/bsw34rzr26clnarkxzgnfqa2la9gx5fy-gcc-4.7.2

$ garden add bj61jjvy9fnm3xxpyq12zpwx2mgg09c8-gcc-4.6.1

$ which gcc
/nix/store/bj61jjvy9f..12zpwx2mgg09c8-gcc-4.6.1/bin/gcc
```

Figure 2) Top pane: Snippet of a garden treetop. Specifies the exact versions to use for all packages in a project. The export list shown bold is used by the *garden add-treetop* command. Bottom pane: typical search and use of a package for scripting.

The Garden provides a solution by a) providing a separate version of all required libraries, up to and including `glibc`, and b) keeping application and library paths unique per version, and c) isolating packages from each other and the OS. No *merge* step is performed; instead an isolation-check of all dynamically linked libraries and executables ensures each dependency also resides in the garden.

Limitations of the Garden include lack of Windows support, and the need to rebuild the package from source: there is limited support for "gardenizing" binary packages. Also no effort is made to block packages that call other binaries by absolute path. While there is some initiative in Ubuntu to adopt NIX as a packaging system [5], there is no built-in support for the garden in common Linux distributions. Finally, the Garden provides support for scripting languages other than Python only through the use of wrapper scripts.

# 3. Architecture

The software garden is a foundation of tools, libraries, and header files for building, organizing, and running software. It allows multiple versions of a package to exist, and provides a robust way to name each version. The garden system ensures all builds contain link and runtime references to other garden packages, i.e. the garden is a closed set under build. The repository of these items is called the "software garden" and is kept in `/garden`, separately from OS versions of the same packages. In the garden:

- A package is kept so long as some version of another package requires it, perhaps forever.
- All packages refer only to other garden packages.
- No package is duplicated.

The garden uses NIX to track packages, establish a clean build environment, and identify dependencies. It also uses NIX's package versioning system.

## 3.1. VERSION

As the Garden supports multiple versions of software, an important question is what is a version? The Garden adopts the concept from NIX. In NIX no assumptions are made regarding what build-time options cause different behavior. The version of gcc 4.5.1 compiled with `--no-java` is different than 4.5.1 compiled without it. The version of gcc built with glibc-2.12.2 is different than when built with glibc-2.12.1, etc.

Garden versions differ from the familiar `MAJOR.MINOR.MICRO` form. They are a hash-name string, with a SHA-256 hash computed from build-time inputs, and a human-readable name and traditional version for convenience. E.g.:

```
bj61jjvy9fnm3xxpyq12zpwx2mgg09c8-gcc-4.6.1
```

The hash inputs are:

- The bytes of the *version.nix* file and garden-helper file (section 3.2.1), which include configure flags used in build, etc.
- Environment variables present at build time.
- System name and architecture (*linux-x86_64*).
- The hash of all dependencies.

The name and final version string are defined by the packager.

## 3.2. USE

The Garden is grown by developers building and pushing packages to a central repository, and users pulling from it. The user-facing command is a shell function called `garden` that alters PATH variables in the environment, with an interface and semantics influenced by the module command (Figure 2) [13].

```
$ garden pull ³
$ garden avail gcc
$ garden show bj61jjpwx2mgg09c8-gcc-4.6.1
$ garden add  bj61jjpwx2mgg09c8-gcc-4.6.1
```

A Python interface is also provided.

```
>>> import garden
>>> garden.add('vxycd107wjbhcj7-numpy-1.5')
>>> import numpy
>>>
```

---

[3] This command rsync's from $GARDEN_CENTRAL. Users may pull an entire garden, a treetop, or a single package.

As long PATH-style variables may affect performance, especially if elements reside on network-mounted directories, the add command ensures multiple adds have a "promotion" effect. If `PATH=A:B:C` and `garden add C` is given, the result is `PATH=C:A:B`. This allows reordering of elements without increasing the list length with redundant entries[4].

The garden environment is setup by a `~/etc/gardenrc` file, sourced by the user's shell, which is initially generated by a bootstrap script.

We considered using the modules system as a frontend to the Garden, but found it was slower and logically cumbersome compared to a simple shell function that called our Python libraries.

### 3.3. BUILD

A Garden package is simply the output of a *make install* into a carefully-versioned output directory. In this section we illustrate how this command is supported: how the build environment is constructed, how packages are checked for proper isolation from the underlying OS, and how they are published for use. To begin building in the garden, the basic GNU toolchain is installed.

Bootstrapping the Garden is a simple matter of installing the NIX tools package, starting the `nix-worker` build daemon, and pulling the NIX-packaged GNU compiler toolchain and common libraries. This is straightforward: installing `gcc`, `boost`, `git`, `python` will populate `/garden` with these tools and their dependencies automatically, in total ~300 pre-compiled packages are pulled from the nix-channel in a few minutes.

#### 3.3.1. VERSION.NIX

In our implementation each product resides in a git repository, which contains a file named `version.nix` that specifies a list of dependencies (Figure 2)[5]. The dependencies are given by a hash-name strings and must name a directory residing in a garden "root" directory. The variable `$GARDEN_STOREPATH` lists the set of garden roots to search. This set containts at least the *public* and *personal* roots. The public root is `/garden` by default and is pointed to by `$GARDEN_ROOT`. The personal root is used for package testing and resides in a developer's home directory. Dependencies are often specified indirectly via the treetop (section 4.2) using simple symbolic names, e.g. `boost`.

A Garden build is ultimately a NIX build; it does not take place in a chroot jail. Instead the build takes places in a highly restricted *environment* with no `PATH` or `HOME` variable set (Figure 4) by a build daemon running as a different user from a git clone of the source. Dependency information is communicated via the environment in variables including `CPPFLAGS` and `LDFLAGS`. This "clean" build in fact occurs rarely; more often a direct build (via the traditional make invocation on the command line) is performed during development. The garden allows direct builds to use the same environment as clean builds.

As the garden is designed for software development, numerous build-test cycles will be performed in rapid succession and must not incur a time penalty. Constructing the build environment takes several seconds due to the design of NIX that requires many `stat` calls to files in the store. Therefore direct builds are preceded by a *configure step* that constructs and saves the build environment to a file. A small wrapper tool called `gmk` (garden make) locates this file and requests `gmk configure` if not found. Else it runs the make command in the clean environment using the `execvpe` call. The speed of a direct build is indistinguishable from a non-garden compile.

The clean build is invoked with the `garden-install` command, and is described further below.

#### 3.3.2. BUILD PATH, CPPFLAGS, LDFLAGS

Most make systems including autoconf will respect the variables CPPFLAGS for finding compile time includes, and LDFLAGS for link instructions [14]. The garden build machinery uses these variables to communicate specific dependencies to a software's unmodified build system[6]. When constructing the build environment certain variables defined in version.nix are treated specially: those named `build_*` pointing to a list of hash-name entries are processed by the garden build. Note these lists generally contain symbolic names that NIX expands to hash-name strings from the treetop (Figure 2).

For example the `build_CPPFLAGS` yields an environment variable `CPPFLAGS` set to its expanded value, using the following logic.

- Locate each hash-name list element using `$GARDEN_STOREPATH`, throwing a clear error if any cannot be resolved. Once package paths have been identified, expect a `$path/garden-env/CPPFLAGS` file containing a list of hash-names.
- Alternatively, for each element check for the existence of a well-known subdirectory, e.g. `$path/include` for `CPPFLAGS`.
- Alternatively, for each element expect a `$path/nix-support/propagated-user-env` file, which contains a list of directories that are added to the result then recursively searched.
- All results are concatenated and joined with a function specific to the task. For CPPFLAGS we join on `-I`. For LDFLAGS we use `-Wl,-rpath` and `-L`, and add `-B binutils/bin` to specify the exact

---

[4] For what a package contributes to PATH see section 3.3.4.
[5] Other SCM systems may be straightforwardly supported.
[6] If a package overwrites or ignores these variables, its make system must be altered.

```
# version.nix
{ storepath, revision }:
let
    VERSION="2.41";
    treetop = "Aien2";
in
package {
    inherit (import treetop storepath) *;
    name = "aien-system-${VERSION}";

    build_CPPFLAGS = [ gcc glibc boost sqlite ]
    build_LDFLAGS  = [ gcc glibc sqlite ]
    build_PATH = [ gcc python stdenv_linux sqlite ];
    build_PYTHONPATH = [ python build_tools ];

    install_command = "./garden-helper --install";
}
```

```
#!/bin/bash
# garden-helper is run in a controlled environment
set -e

case $1 in
  --install)
    configure -prefix=$out
    make
    make install
    mkdir -p $out/garden-env
    echo "$out/bin" > $out/garden-env/PATH
    echo "$out/lib64" > $out/garden-env/LDFLAGS
    echo "$out/include > $out/garden-env/CPPFLAGS
    echo "$out/lib-python" > $out/garden-env/PYTHONPATH
    ;;
esac
```

Figure 3) A package's garden makefiles. The **version.nix** in top pane specifies the product's dependencies; the **garden-helper** script in the bottom pane controls the build and defines how to compose itself in other software. Both are stored with the package's source code.

```
TMP=/tmp/nix-build-hfc1gi43crfv44dp1ih6jg00ywqf9lsn-
aien-trading-system-2.41.drv-0
HOME=/homeless-shelter
PATH=/path-not-set
gcc=/nix/store/bj61jjvy9fnm3xxpyq12zpwx2mgg09c8-gcc-
4.6.1
out=/nix/store/gncj2dy55v6nicybra92vb5k76yiqa12-aien-
trading-system-2.41
system=x86_64-linux
install_command="./garden-helper --install"
CPPFLAGS=-I/nix/store/xxx-gcc-4.7.2 -I/nix/store/sdf-
glibc-2.13 -I/nix/store/yyy-boost-1.51
LDFLAGS=-L/nix/store/xxx-gcc-4.7.2 -Wl,-
rpath/nix/store/xxx-gcc-4.7.2
```

```
$ ldd build/toy_app
  libboost_program_options.so.1.51.0 =>
/nix/store/4bsw2ldgn170swma9hsif7hs61qdv86g-boost-
1.51.0/lib/libboost_program_options.so.1.51.0
  libstdc++.so.6 =>
/nix/store/bsw34rzr26clnarkxzgnfqa2la9gx5fy-gcc-
4.7.2/lib64/libstdc++.so.6
  libm.so.6 =>
/nix/store/cj7a81wsm1ijwwpkks3725661h3263p5-glibc-
2.13/lib/libm.so.6
/nix/store/cj7a81wsm1ijwwpkks3725661h3263p5-glibc-
2.13/lib/ld-linux-x86-64.so.2
```

Figure 4) Top pane: Snippet of the build environment. No PATH nor HOME nor USER is set. The **out** variable is the final location of the build.

Bottom pane: isolation check. All libraries are linked with -rpath, ensuring all runtime dependencies are found in the garden, including the loader (dynamic linker, bold).

assembler and linker, and `-B glibc/lib` to find the C-runtime libraries. We also give `--dynamic-linker glibc/lib` to fully isolate the executable. The set of recognized build names currently support C/C++ compilation, but are extensible in a Python library.

An `ldd` of a correctly-made executable in the Garden will show only lines containing `=> /garden`, i.e. all dependencies reside in the Garden. This test is automated in garden-helpers by the `garden-check-ldd-clean` tool, which aborts the build on failure:

```
for i in $(find -name "*.so")
do
 garden-check-ldd-clean $i
done
```

In this way, unclean garden packages have a reasonable chance of being caught before published.

### 3.3.3. GARDEN-INSTALL

The garden-install command performs the configure, build, and install steps with various levels of paranoia, and then may send the built package to a garden repository another box via rsync.

```
$ garden-install --personal
$ garden-install git:HEAD
$ garden-install git:HEAD --push
$ garden-install git:HEAD --export
```

The `garden-install` tool follows the same procedure as a direct build, but takes harsher steps to ensure the result is repeatable. First it refuses to operate if any local changes have not been checked into source control. Second, it takes a `git:revspec` argument and makes a clean checkout (git local clone) of source into a temp directory. Third it sends the nix-worker daemon a build request via a local socket. This worker, running as a neutral user, sets up the build environment and calls the package's `garden-helper` script (Figure 3). This command must perform the "make install" to populate the package into the `$out` directory, and create the composition instructions (see next section). The package is aborted if the helper script exits with an error or `$out` is unpopulated.

The first three steps are skipped if the `--personal` flag is given, to allow a user to test the garden-helper logic. In this case the garden-helper is invoked as the user running the build, from the possibly dirty working directory. The install target is `$GARDEN_PERSONAL_ROOT`, and the resulting package can be used in other personal installs. A package is often installed into the personal root for inspection. Only

after the developer is completely satisfied with the packaging is it built in the public garden and exported to the central garden root for users to pull[7].

### 3.3.4. PACKAGE COMPOSITION

Each package in the Garden is responsible for instructing other packages on its use, for example where to find its include files, libraries, and binaries. These instructions are stored in the `garden-env/` directory of a package, in files named after the build variables they affect. E.g. the contents of the file `$out/garden-env/CPPFLAGS` is used to inject this package's includes into the CPPFLAGS variable of any package that uses it. The files may contain complier flags, and all hash-name strings are searched for and expanded to full garden paths. While this mechanism may sound complex, it is often simple to construct:

```
hashname=$(basename $out)
echo "$hashname/bin" > $out/garden-env/PATH
echo "$hashname/lib" > $out/garden-env/LDFLAGS
```

Some packages may require an environment variable set before use, for example to specify a license server. To provide this a package creates a file at build-time that is sourced by the shell during `garden add`. In this way a package may specify arbitrary initialization logic:

```
garden add v6ha68vx-icc-4.0⁸
+source $GARDEN_ROOT/v6ha68vx-icc-4.0/garden-env/default.sh
+export ICC_LICENSE_SERVER=license-server1
```

Another requirement of *add* may be to pre-load another package at runtime; for example the Python module `scipy` requires a particular `numpy`. This is achieved by a similar structure, `$out/garden-env/DEPS/`, that contains files listing packages to pre-load, named after the variables they effect. For example the file `$scipy/garden-env/DEPS/PYTHONPATH` contains the hash-name of the `numpy` it was built and tested with. The add mechanism avoids unchecked recursion by checking for circular dependencies.

### 3.4. EXPORT AND PRODUCTION

The purpose of export is to build the package and transfer it and all its runtime dependencies to a remote production machine or central repository. This machine is assumed to have no source code on local or available disks.

The export process (Figure 5) begins with a public build as described above, and the final package path is recorded. The export logic uses *version.nix* to find all named dependencies of the package, and for each one queries its dependencies recursively (with `nix-store -qR`). The freshly-installed product is added to this list, then all are synced to the host and path specified with $GARDEN_CENTRAL_DEST using `rsync`. While this list can be long, much of it already exists on the destination, and rsync ensures each package is sent only once. For the aien system the export closure is 175 packages. As an optimization, the `--push` option only sends the single newly-built package. Finally a multicast packet is emitted to alert other garden repositories a new package is available. This signal is currently ignored, but could activate a garden pull or other mechanism to bring all gardens into consistency.

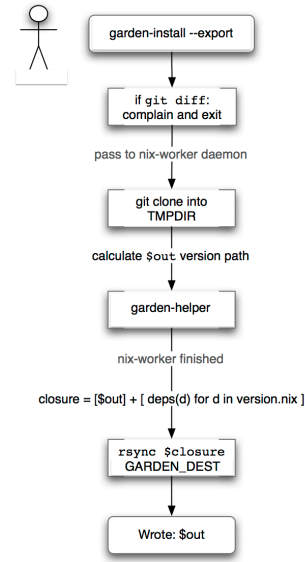

Figure 5) Export procedure. The process ensures the published package includes all intended changes, and builds from a repeatable source point.

### 3.4.1. RUN

In our aien-system example the software package is a long-running daemon controlled by an init script. Activating the software is a simple matter of editing this script with the path to the new daemon executable and restarting the service. The more general case of using a package is shown in section 3.2.

### 3.4.2. ROLLBACK

As multiple versions exist (ideally in perpetuity), a rollback of a software package such as Python involves nothing more than selecting the desired previous version and running `garden add`. For a long-running daemon, the initscript is similarly edited with the previous version and the service restarted.

A more difficult problem arises if the daemon relies on a configuration file whose schema has changed. The garden only contains software, not generations of data; a previous version of the configuration must be separately kept and identified with a garden version.

---

[7] When installing in the public store, no personal dependencies are allowed.
[8] A `set +x` was added for illustration.

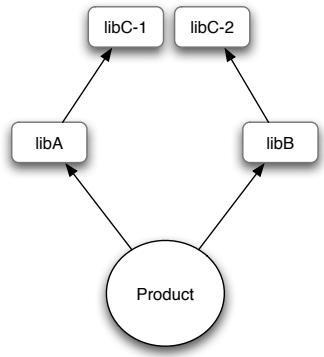

Figure 6) Diamond dependency problem. Product cannot link as the linker cannot handle two versions of *libC*. Resolved by building each component with the same treetop.

For libraries, rollback generally involves all clients being rebuilt with a newer version, and simply leaving the offending library in place. It is rare to remove a malfunctioning library version from the garden if there exists at least one correctly functioning client, or practically one relied-upon client that for various reasons is difficult to rebuild.

### 3.4.3. COMPOSING GARDENS

Multiple garden repositories may exist in the same HPC installation. It may be necessary to keep one more tightly access-controlled than another for example. From the user's view, multiple gardens involves simply adding the `GARDEN_ROOT` of each to `$GARDEN_STOREPATH` and the new packages will be visible to `garden avail`. The same variable is used by developers for build. Failure to do so will be graceful, e.g. if one builds a package which requires a library or build tool in an inaccessible garden a clear error with the full hash-name(s) will be printed.

Due to linking with rpath and absolute paths, executables expect to find libraries in a well-known location, i.e. `/garden`. This stable prefix makes composing multiple gardens possible. While garden builds could use the $ORIGIN feature of the linker to locate the garden root and so support garden relocation, the variable would break executables with cross-links between gardens. For the $ORIGIN technique to work, all gardens would have to be in the same location on developer's workstations as on a user's HPC clusters. This is neither enforceable nor convenient.

Therefore we compose gardens by locating garden roots wherever practical, and symlinking all packages so they are available in `/garden`. This may be done for each new package with the command

```
ln -s $GARDEN_ROOT/store/$path /garden
```

The result is well defined by the hash-name semantics: if two packages have the same hash, they are the same package. While imposing additional logic on the garden pull command, this mechanism ensures the loader finds all available libraries no matter where gardens reside.

### 3.4.4. EXPERIMENTAL VERSIONS

Clearly one benefit of the Garden is the ease with which development, debug, and experimental versions of products can be built and tested. The Garden relieves the concern these alternate builds will break existing, well-tested software, as all versions co-exist independently. In addition the NIX structure naturally de-duplicates common libraries, making export fast and reducing the disk space footprint of many experimental versions on hosts.

## 4. Issues and Usage

In this section we examine some subtleties and difficulties when using the Garden.

### 4.1. CLIENTS OF RUNNING SERVICES

While the Garden provides great flexibility for developers and stability for users, one limitation is client software to interface with a long-running service. The client version must be matched to the service version, and scripts may "just break" when the service is upgraded due to API incompatibilities. For example while users loved and trusted `slurm-1.2` client libraries for submitting HPC jobs, they must change when the server is upgraded. A similar problem exists when interfacing with kernel modules. Generally this problem is solved by maintaining a small system-wide garden-init file in `/etc/profile.d/` kept up to date by a configuration management system.

### 4.2. TREETOPS AND DIAMOND DEPENDANCIES

Multiple versions can cause a "diamond dependency" problem where a product indirectly requires multiple versions of the same library (Figure 6). Such a situation is surprisingly easy to achieve as the codebase grows, and results in impossible requests to the linker.

A solution is to grow a clique of libraries that have no diamond dependencies, and build the product using only libraries from this set. A *treetop* serves as a common "include file" for all libraries used in a project (Figure 6). It names the exact versions of each tool and library; if a non-leaf package is updated (e.g. glibc or boost), all dependant packages are rebuilt. While automated rebuild is straightforward, it is currently a manual process for Garden software due to our finding that automated testing is inadequate in many packages. Treetops are versioned like other packages in the garden, or maintained by git. A product's version.nix file names the treetop it will use (Figure 3).

### 4.3. CHOOSING AND PRUNING

To use a garden package one adds it with `garden add hash-name` in shell or Python. As the Garden does not link "default" versions into a well-known location such as

`/usr/bin`, users enumerate available packages with `garden avail` and choose one. Which to choose is of course is the crux of using the garden; that choice is outside the scope of the system. It is often a version you have been told about in an email, wiki page, or one taken from another script. The garden maintains the "born-on" date of each package, as well as the git revision a public install was built from to aid the choice.

To allow operators to conveniently define an environment, a treetop can specify a set of packages to add together, perhaps to set the project-chosen Python and GCC tools, and the Infiniband library version known to work with currently-available hardware. The `garden add-treetop` performs this operation, adding all the packages from a treetop's *export* list.

Once chosen, these add commands may be placed in a user's shell configuration, or may be used in the header of a shell script, e.g. to set a particular version of Python for that script. That latter usage ensures correct behavior even if another user (with a different *.bashrc*) obtains and runs the script.

A related problem is pruning old and unused packages from the garden to reclaim disk resources. After several years our garden grew to a few hundred Gigabytes in size. However it is difficult to predict which packages truly are unused, as rollback may require a long-unused version. A complete solution is left for future work. One policy may be to delete packages not currently used but record their git revision and repository URL; if a deleted version is needed for rollback it may be simply recreated.

### 4.4.  3$^{RD}$ PARTY SOFTWARE

Software obtained from a third party may be placed into the garden even if it was not designed for such installation. The procedure is similar to that described in section 3.3.3; the packager constructs a version.nix with all required dependencies (often from a treetop) and a garden-helper to guide the untar, build, and install phases, as well as to create the package composition instructions.

If the package follows the GNU build standards [14,15] this process is straightforward. If build variables such as LDFLAGS are ignored or overridden, a patch to its build system must be applied by garden-helper.

### 4.5.  POLICING CLEAN BUILDS

There is no system in the Garden to ensure a build maintains the closure property, i.e. that all its library references are within the garden[9]. Unlike CDE, we make no examination of the runtime behavior during the packaging step, instead relying on the build script to do this work. Especially when building 3$^{rd}$-party software that uses Autoconf, it is not straightforward to ensure the package is truly isolated from OS-provided packages; autoconf's `configure` script has a strong affinity to the `/usr/lib` location when searching for dependencies. However good garden-helper practices, in particular careful use of the ldd parser described in section 3.3.2 can help.

## 5.  Conclusion

In this paper we presented the design and architecture of the Software Garden, a method of building custom software that supports and exposes all versions of the software package ever published, at every stage of development and operation. By exactly specifying versions of build tools and runtime dependencies, all developers may produce the same output despite slight differences in their workstation OS, and old versions of a software package may be trivially and precisely recreated even long after they have been replaced in production.[10] It also provides a well-defined base for product deployment in a customer's environment.

Because each iteration of software is stored in a uniquely named location, the Garden ensures installation will never affect currently running software; an important feature in 24/7 environments and one shared with NIX. Unlike NIX, the Garden is suited for custom software development, and not simply software packaging; furthermore it works safely with existing OS distributions and their packaging systems. The Garden makes deterministic rollback of a single package possible, without affecting the behavior of other packages. This practical ability allows dead-of-night rollbacks by an operations team without coordination with developers, leading to less stress and more sleep by all.

Due to its built-time isolation checks that ensure garden packages refer only to other garden packages, the Garden protects production software from routine upgrades to the underlying OS. It has no performance penalty over standard build techniques as it adds no additional layers or runtime complexity; the mechanisms used to ensure software isolation already exist in the GNU build toolchain.

For architects, the Garden makes exploring new libraries and compiler toolchains easy – one developer may cut a version that tries a new tool and if successful can run it in production without affecting established product versions on the system. The Garden makes trying new APIs easy for the same reason; one library version may implement API improvements without forcing all clients to upgrade. Only new clients who *specifically request it* will be affected, a useful property that quickly made the system indispensable to developers. The treetop concept helps development as well by restricting choice to ensure all software in a project links with the same version of base libraries.

---

[9] For C/C++ these are shared objects accessed by the loader and *dlopen* calls, for Python these are modules opened by *import*.

[10] This has important benefits, e.g. for questions of scientific reproducability, and for legal compliance requirements.

The garden is naturally suited to C/C++ and other compiled language projects; support for the Python scripting languages is also included. We illustrated the practical development and operations lifecycle of a long-running HPC system, which has employed the Garden for over a year in full production.

## 5.1. ACKNOWLEDGEMENTS

We would like to thank Kevin Bowers for the initial germ of a Garden (and the name); and John Salmon and Mark Moraes for developing this germ into a non-NIX Garden system. Ross Lippert for the treetop concept, and Eelco Dolstra for his help with NIX. Thanks also to David Potter and Sebastien Donadio for feedback on early versions of the paper.